\documentclass[runningheads]{llncs}

\usepackage[T1]{fontenc}
\usepackage{hyperref}

\def\doi#1{\href{https://doi.org/\detokenize{#1}}{\url{https://doi.org/\detokenize{#1}}}}
\usepackage{graphicx}
\usepackage{listings}
\begin{document}
\title{Edge-aware Multi-task Network for Integrating Quantification Segmentation and Uncertainty Prediction of Liver Tumor on Multi-modality Non-contrast MRI}
% Integrating 
\author{Xiaojiao Xiao\inst{1}\and Qinmin Hu\inst{1} \and Guanghui Wang\inst{1} \thanks{Corresponding author: Guanghui Wang ({wangcs@torontomu.ca})}}
%

\institute{Department of Computer Science, Toronto Metropolitan University, Toronto, Canada}

\maketitle              % typeset the header of the contribution
\begin{abstract}
Simultaneous multi-index quantification, segmentation, and uncertainty estimation of liver tumors on multi-modality non-contrast magnetic resonance imaging (NCMRI) are crucial for accurate diagnosis. However, existing methods lack an effective mechanism for multi-modality NCMRI fusion and accurate boundary information capture, making these tasks challenging. To address these issues, this paper proposes a unified framework, namely edge-aware multi-task network (EaMtNet), to associate multi-index quantification, segmentation, and uncertainty of liver tumors on the multi-modality NCMRI. The EaMtNet employs two parallel CNN encoders and the Sobel filters to extract local features and edge maps, respectively. The newly designed edge-aware feature aggregation module (EaFA) is used for feature fusion and selection, making the network edge-aware by capturing long-range dependency between feature and edge maps. Multi-tasking leverages prediction discrepancy to estimate uncertainty and improve segmentation and quantification performance. Extensive experiments are performed on multi-modality NCMRI with 250 clinical subjects. The proposed model outperforms the state-of-the-art by a large margin, achieving a dice similarity coefficient of 90.01$\pm$1.23 and a mean absolute error of 2.72$\pm$0.58 mm for MD. The results demonstrate the potential of EaMtNet as a reliable clinical-aided tool for medical image analysis.

\end{abstract}

\section{Introduction}
\begin{figure}[t]
\includegraphics[width=\textwidth]{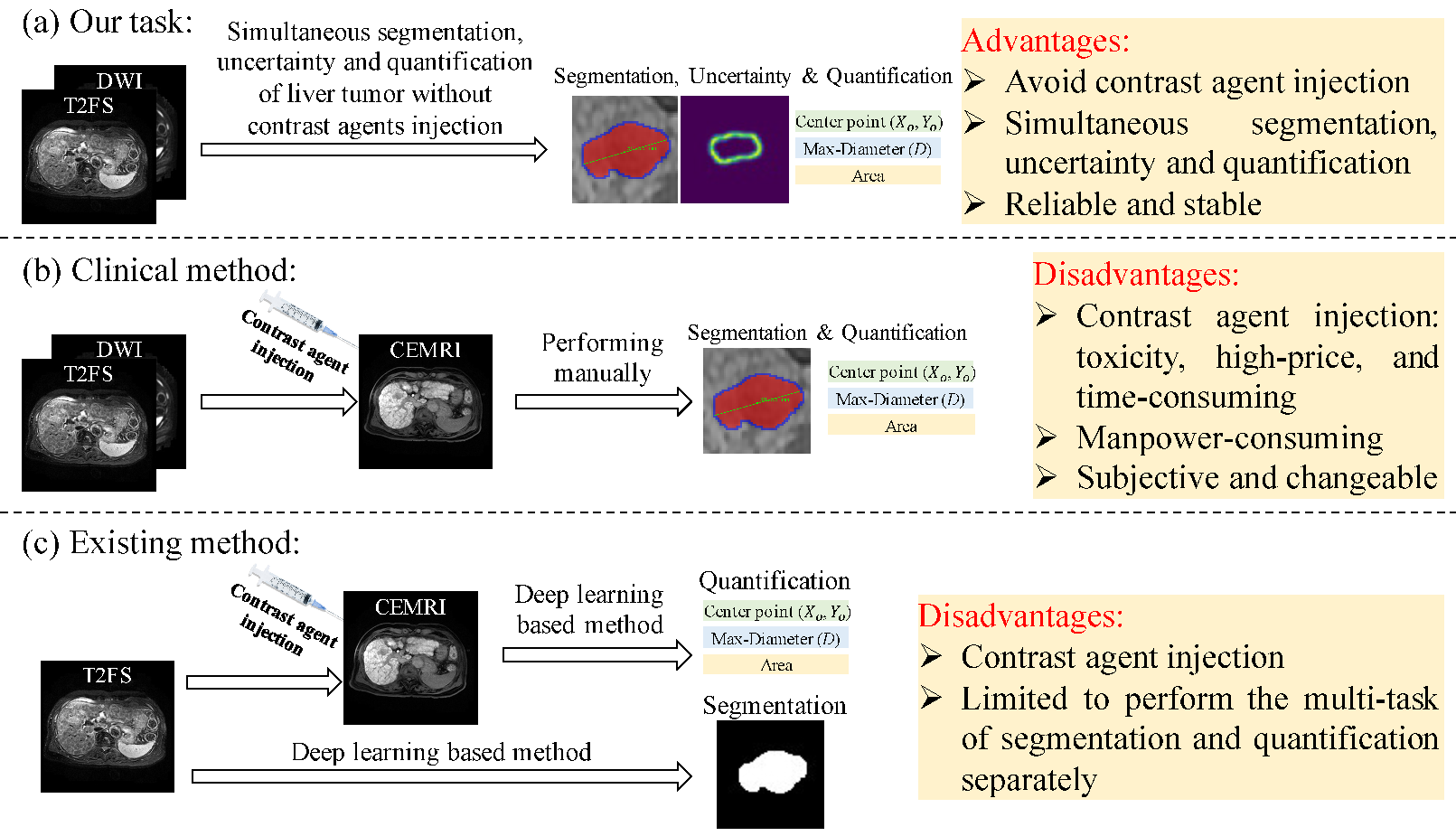}
\caption{Our method integrates segmentation and quantification of liver tumor using multi-modality NCMRI, which has the advantages of avoiding contrast agent injection, mutual promotion of multi-task, and reliability and stability. } \label{figure1}
\end{figure}

Simultaneous multi-index quantification (i.e., max diameter (MD), center point coordinates ($X_{o}$, $Y_{o}$), and Area), segmentation, and uncertainty prediction of liver tumor have essential significance for the prognosis and treatment of patients \cite{gonzalez2013,rovira2008}. In clinical settings, segmentation and quantitation are manually performed by the clinicians through visually analyzing the contrast-enhanced MRI images (CEMRI) \cite{huo2018supervoxel,lee2015hepatocellular,sirlin2014consensus}. However, as shown in Fig.\ref{figure1}(b), Contrast-enhanced MRI (CEMRI) has the drawbacks of being toxic, expensive, and time-consuming due to the need for contrast agents (CA) to be injected \cite{danet2003,fishbein2005}. Moreover, manually annotating medical images is a laborious and tedious process that requires human expertise, making it manpower-intensive, subjective, and prone to variation \cite{petitclerc2017}. Therefore, it is desirable to provide a reliable and stable tool for simultaneous segmentation, quantification, and uncertainty analysis, without requiring the use of contrast agents, as shown in Fig.\ref{figure1}(a).

Recently, an increasing number of works have been attempted on liver tumor segmentation or quantification \cite{xiao2022task,xiao2019,zhang2021weakly,zhao2021mftrans}. As shown in Fig.\ref{figure1} (c), the work \cite{xiao2019} attempted to use the T2FS for liver tumor segmentation, while it ignored the complementary information between multi-modality NCMRI of T2FS and DWI. In particular, there is evidence that diffusion-weighted imaging (DWI) helps to improve the detection sensitivity of focal lesions as these lesions typically have higher cell density and microstructure heterogeneity \cite{tang2019diffusion}. The study in \cite{xiao2022task,zhao2021mftrans} attempted to quantify the multi-index of liver tumor, however, the approach is limited to using multi-phase CEMRI that requires the injection of CA. In addition, all these works are limited to a single task and ignore the constraints and mutual promotion between multi-tasks. Available evidence suggests that uncertainty information regarding segmentation results is important as it guides clinical decisions and helps understand the reliability of the provided segmentation. However, current research on liver tumors tends to overlook this vital task.

To the best of our knowledge, although many works focus on the simultaneous quantization, segmentation, and uncertainty in medical images (i.e., heart \cite{du2018direct,ge2019k,luo2020commensal,xu2020segmentation}, kidney \cite{ruan2020mb}, polyp \cite{patel22fuzzynet}). No attempt has been made to automatically liver tumor multi-task via integrating multi-modality NCMRI due to the following challenges: (1) The lack of an effective multi-modality MRI fusion mechanism. Because the imaging characteristics between T2FS and DWI have significant differences (i.e., T2FS is good at anatomy structure information while DWI is good at location information of lesions \cite{zhao2021united}). (2) The lack of strategy for capturing the accurate boundary information of liver tumors. Due to the lack of contrast agent injection, the boundary of the lesion may appear blurred or even invisible in a single NCMRI, making it challenging to accurately capture tumor boundaries \cite{zhao2021united}. (3) The lack of an associated multi-task framework. Because segmentation and uncertainty involve pixel-level classification, whereas quantification tasks involve image-level regression \cite{luo2020commensal}. This makes it challenging to integrate and optimize the complementary information between multi-tasks.

In this study, we propose an edge-aware multi-task network (EaMtNet) that integrates the multi-index quantification (i.e., center point, max-diameter (MD), and Area), segmentation, and uncertainty. Our basic assumption is that the model should capture the long-range dependency of features between multi-modality and enhance the boundary information for quantification, segmentation, and uncertainty of liver tumors. The two parallel CNN encoders first extract local feature maps of multi-modality NCMRI. Meanwhile, to enhance the weight of tumor boundary information, the Sobel filters are employed to extract edge maps that are fed into edge-aware feature aggregation (EaFA) as prior knowledge. Then, the EaFA module is designed to select and fuse the information of multi-modality, making our EaMtNet edge-aware by capturing the long-range dependency of features maps and edge maps. Lastly,  the proposed method estimates segmentation, uncertainty prediction, and multi-index quantification simultaneously by combining multi-task and cross-task joint loss.
 
The contributions of this work mainly include: (1) For the first time, multi-index quantification, segmentation, and uncertainty of the liver tumor on multi-modality NCMRI are achieved simultaneously, providing a time-saving, reliable, and stable clinical tool. (2) The edge information extracted by the Sobel filter enhances the weight of the tumor boundary by connecting the local feature as prior knowledge. (3) The novel EaFA module makes our EaMtNet edge-aware by capturing the long-range dependency of features maps and edge maps for feature fusion. The source code will be available on the author's website.

\section{Method}
\begin{figure}[t]
\includegraphics[width=\textwidth]{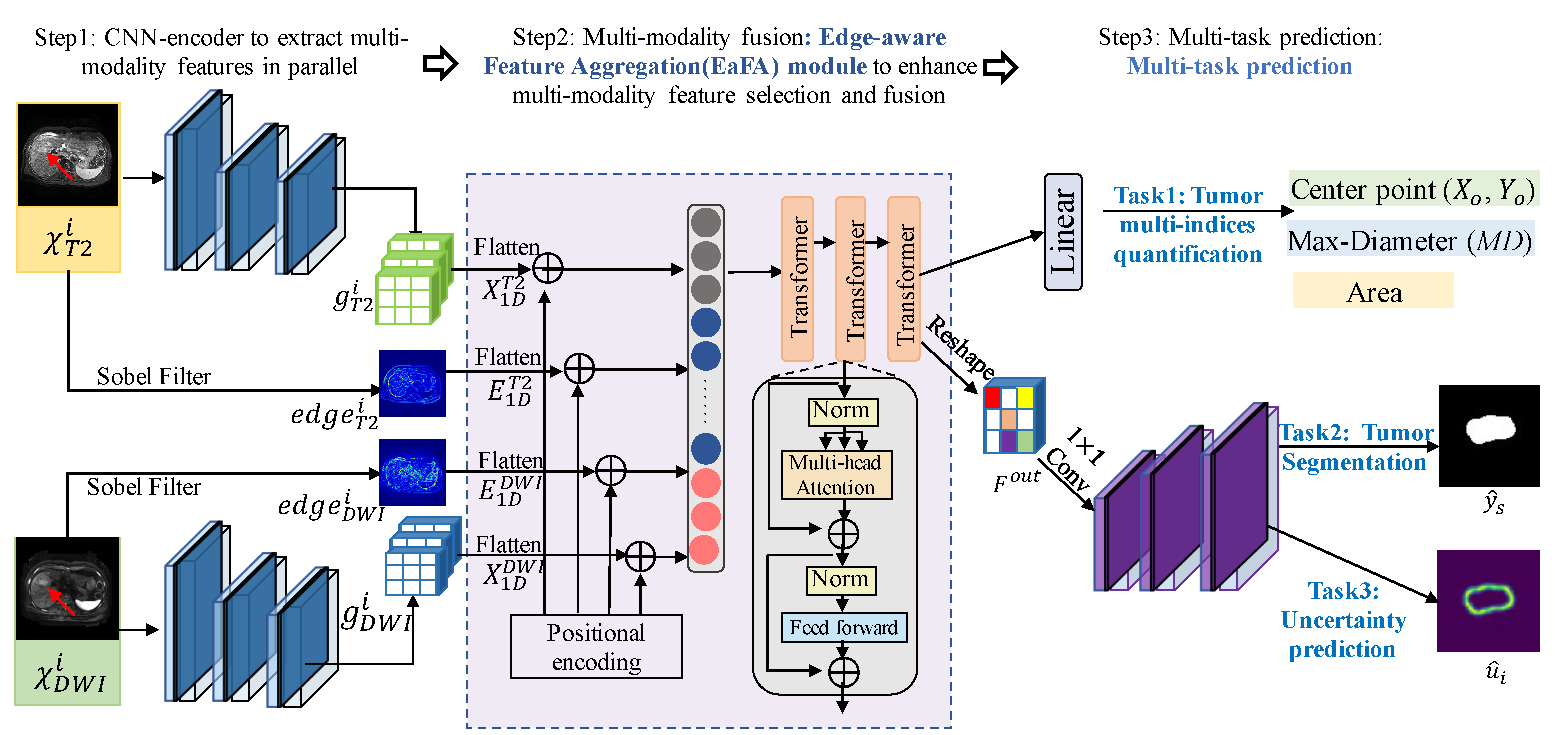}
\caption{Overview of the EaMtNet. It mainly consists of three steps: 1) CNN encoders for extracting local features of multi-modality NCMRI; 2) EaFA for enhancing multi-modality NCMRI feature selection and fusion; and 3) Multi-task prediction.} \label{figure2}
\end{figure}

The EaMtNet employs an innovative approach for simultaneous tumor multi-index quantification, segmentation, and uncertainty prediction on multi-modality NCMRI. As shown in Fig.\ref{figure2}, the EaMtNet inputs multi-modality NCMRI of T2FS and DWI for capturing the feature and outputs the multi-index quantification, segmentation, and uncertainty. Specifically, the proposed approach mainly consists of three steps: 1) The CNN encoders for capturing feature maps and the Sobel filters for extracting edge maps (Section 2.1); 2) The edge-aware feature aggregation (EaFA) for multi-modality feature selection and fusion via capturing the long-distance dependence (Section 2.2); and 3) Multi-task prediction module (Section 2.3).

\subsection{CNN encoder for feature extraction}
In Step 1 of Fig. \ref{figure2}, the multi-modality NCMRI (i.e., $\chi_{T2}^{i}\in R^{H\times W}$, $\chi _{DWI}^{i}\in R ^{H\times W}$) are fed into two parallel encoders and the Sobel filter to extract the feature maps (i.e., $g _{T2}^{i} \in R ^{H\times W\times N} $, $g _{DWI}^{i}\in R ^{H\times W \times N}$) and the corresponding edge maps (i.e., $edge _{T2}^{i}\in R ^{H\times W}$, $edge _{DWI}^{i}\in R ^{H\times W}$) respectively. Specifically, EaMtNet employs UNet as the backbone for segmentation because the CNN encoder has excellent capabilities in low-range semantic information extraction \cite{ronneberger2015u}. The two parallel CNN encoders have the same architecture where each encoder contains three shallow convolutional network blocks to capture features of adjacent slices. Each conv block consists of a convolutional layer, batch normalization, ReLU, and non-overlapping subsampling. At the same time, EaMtNet utilizes the boundary information extracted by the Sobel filter \cite{sobel19683x3} as prior knowledge to enhance the weight of tumor edge information to increase the awareness of the boundary.

\subsection{Edge-aware feature aggregation(EaFA) for multi-modality feature selection and fusion}
In Step 2 of the proposed model, the feature maps (i.e., $g _{T2}^{i}$, $g _{DWI}^{i}$) and the edge maps (i.e., $edge_{T2}^{i}$, $edge_{DWI}^{i}$) are fed into EaFA for multi-modality feature fusion with edge-aware. In particular, the EaFA makes the EaMtNet edge-aware by using the Transformer to capture the long-range dependency of feature maps and edge maps. Specifically, the feature maps and edge maps are first flattened to the 1D sequence corresponding to $X_{1D}\in R^{N\times P^2}$ and $E_{1D}\in R^{2\times Q^2}$, respectively. Where $N= 2\times C$ means the channel number C of the last convolutional layer from the two parallel encoders. $(P, P)$ and $(Q, Q)$ represent the resolution of each feature map and each edge map, respectively. On the basis of the 1D sequence, to make the feature fusion with edge awareness, the operation of position encoding is performed not only on feature maps but also on edge maps. The yielded embeddings $Z\in R^{N\times P^2+2\times Q^2}$ can serve as the input sequence length for the multi-head attention layer in Transformer. The following operations in our EaFA are similar to the traditional Transformer \cite{vaswani2017attention}. After the three cascade Transformer layers, the EaFA yields the fusion feature vector $F$ for multi-task prediction. The specific computation of the self-attention matrix and multi-head attention are defined below \cite{vaswani2017attention}:
\begin{equation}
Attention(\mathcal{Q}, \mathcal{K}, \mathcal{V})=softmax(\frac{\mathcal{Q}\mathcal{K}^\mathcal{T}}{\sqrt{d_k}})(\mathcal{V}) \\
\end{equation}
\begin{equation}
MultiHead(\mathcal{Q}, \mathcal{K}, \mathcal{V})=Concat(head_1,...,head_h)\mathcal{W}^\mathcal{O}
\end{equation}
\begin{equation}
head_i=Attention(\mathcal{Q}\mathcal{W}^\mathcal{Q}_i, \mathcal{K}\mathcal{W}^\mathcal{O}_i, \mathcal{V}\mathcal{W}^\mathcal{V}_i)
\end{equation}
where query $\mathcal{Q}$, key $\mathcal{K}$, and value $\mathcal{V}$ are all vectors of the flattened 1D sequences of $X_{1D}$ and $E_{1D}$. $\mathcal{W}^\mathcal{O}_i$ is the projection matrix, and $\frac{1}{\sqrt{d_k}}$ is the scaling factor.

\subsection{Multi-task prediction}
In Step 3 of Fig. \ref{figure2},  the EaMtNet outputs the multi-modality quantification $\hat{y}_{Q}$ (i.e., MD, $X_{o}$, $Y_{o}$ and Area), segmentation result $\hat{y}_{s}$ and uncertainty map $\hat{u}_{i}$.  Specifically, for the quantification path, $\hat{y}_{Q}$ is directly obtained by performing a linear layer to the feature $F$ from EaFA. For the segmentation and uncertainty path, the output feature $F$ from EaFA is first reshaped into a 2D feature map $F^ {out}$. Then, to scale up to higher-resolution images, a $1\times 1$ convolution layer is employed to change the channel number of $F^ {out} $ for feeding into the decoder. After upsampling by the CNN decoder, EaMtNet predicts the segmentation result $\hat{y}_{s}$ with $H\times W$ and uncertainty map $\hat{u}_{i}$ with $H\times W$. The CNN decoder contains three shallow deconv blocks, which consist of deconv layer, batch normalization, and ReLU. Inspired by \cite{wang2017gated}, we select the entropy map as our uncertainty measure. Given the prediction probability after softmax, the entropy map is computed as follows:
\begin{equation}
H[x]=-\sum_{i=1}^{K}z_{i}(x)\ast log_{2}(z_{i}(x)) \\
\end{equation}
where $z_{i}$ is the probability of pixel $x$ belonging to category $i$. When a pixel has high entropy, it means that the network is uncertain about its classification. Therefore, pixels with high entropy are more likely to be misclassified. In other words, its entropy will decrease when the network is confident in a pixel's label. 

Under the constraints of uncertainty, the EaMtNet can effectively rectify the errors in tumor segmentation because the uncertainty estimation can avoid overconfidence and erroneous quantification \cite{wang2019aleatoric}. Moreover, the EaMtNet novelly make represent different tasks in a unified framework, leading to beneficial interactions. Thus, the quantification performance is improved through back-propagation by the joint loss function $L_{multi-task}$.  The function comprises segmentation loss $L_{seg}$ and quantification loss $L_{qua}$, where the loss function $L_{seg}$ is utilized for optimizing tumor segmentation, and $L_{qua}$ is utilized for optimization of multi-index quantification. It can be defined as:
\begin{equation}
{L_{Dice}}=\frac{2\sum_{i}^{N} y_{i} \hat{y}_{s}}{\sum_{i}^{S}{ y_{i}}^{2}+\sum_{i}^{N}{\hat{y}_{s}}^{2}}
\end{equation}
 \begin{equation}  
 L_{qua}\left (  \hat{y}_{task}^{i}, {y}_{task}^{i} \right )=\sum_{i=1}\left| {{y}_{task}^{i}}- {\hat{y}_{task}^{i}} \right |
 \end{equation}	
 where $\hat{y}_{s}$ represents the prediction, and $y_{i}$ represents the ground truth label. The sum is performed on $S$ pixels, $\hat{y}_{task}^{i}$ represents the predicted multi-index value, and ${y}_{task}^{i}$ represents the ground truth of multi-index value,  ${task}$ $\in$ \{$MD$, $X$, $Y$, $Area$\}.

 \section{Experimental results and discussion}
 
For the first time, EaMtNet has achieved high performance with the dice similarity coefficient (DSC) up to 90.01$\pm$1.23\%, and the mean absolute error (MAE) of the MD, $X_{o}$, $Y_{o}$ and Area are down to 2.72$\pm$0.58 mm,1.87$\pm$0.76 mm, 2.14$\pm$0.93 mm and 15.76$\pm$8.02 $cm^2$, respectively.
 
\subsubsection{Dataset and configuration.} An axial dataset includes 250 distinct subjects, each underwent initial standard clinical liver MRI protocol examinations with corresponding pre-contrast images (T2FS [4mm]) and DWI [4mm]) was collected. The ground truth was reviewed by two abdominal radiologists with 10 and 22 years of experience in liver imaging, respectively. If any interpretations demonstrated discrepancies between the reviewers, they would re-evaluate the examinations together and reach a consensus. To align the paired images of T2 and DWI produced at different times. We set the T2 as the target image and the DWI as the source image to perform the pre-processing of non-rigid registration between T2 and DWI by using the Demons non-rigid registration method. It has been widely used in the field of medical image registration since it was proposed by Thirion \cite{thirion1996non}. We perform the Demons non-rigid registration on an open-source toolbox DIRART using Matlab 2017b.

Inspired by the work \cite{vaswani2017attention}, we set the scaling factor $d_k$ to 64 in equation (1). All experiments were assessed with a 5-fold cross-validation test. To quantitatively evaluate the segmentation results, we calculated the dice coefficient scores (DSC) metric that measures the overlapping between the segmentation prediction and ground truth \cite{milletari2016v}. To quantitatively evaluate the quantification results, we calculated the mean absolute error (MAE). Our EaMtNet was implemented using Ubuntu 18.04 platform, Python v3.6, PyTorch v0.4.0, and running on two NVIDIA GTX 3090Ti GPUs. 

\begin{figure}[t]
\includegraphics[width=\textwidth]{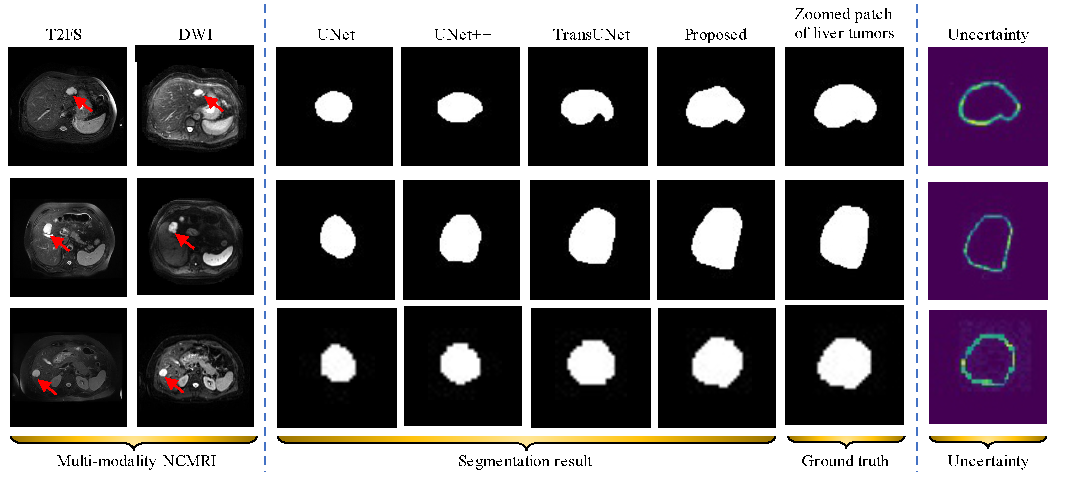}
\caption{The comparison of segmentation results between the proposed EaMtNet and three SOTA methods. The results show that our network yields high performance.} \label{figure3}
\end{figure}
 \begin{table}[t]
	\center{}
	\caption{Segmentation and quantification performance of the EaMtNet under different configurations. DSC is used to evaluate the segmentation performance. MAE is used to evaluate the quantification performance.
	}\label{tab1}
	\begin{tabular}{|l|l|l|l|l|l|l|l}
		\hline
		& DSC(\%)&MD($mm$)&$X_{o}$($mm$)&$Y_{o}$($mm$)&Area($cm^2$)\\
		\hline				        
		No-EaFA& 85.82$\pm$1.92&3.49$\pm$0.94&2.51$\pm$1.43&3.12$\pm$1.84&28.55$\pm$9.75\\    No-Uncertainty&88.37$\pm$2.71&3.25$\pm$0.77&2.36$\pm$0.92&2.78$\pm$1.18&24.15$\pm$9.19\\	   		
		Our method & 90.01$\pm$1.23&2.72$\pm$0.58&1.87$\pm$0.76&2.14$\pm$0.93&15.76$\pm$8.02\\
		                    		\hline
	\end{tabular}
\end{table}

\subsubsection{Accurate segmentation.} 
The segmentation performance of EaMtNet has been validated and compared with three state-of-the-art (SOTA) segmentation methods (TransUNet \cite{chen2021transunet}, UNet \cite{ronneberger2015u}, and UNet++ \cite{zhou2019unet++}). Furthermore,  to ensure consistency in input modality, the channel number of the first convolution layer in the three comparison methods is set to 2. The visual examples of liver tumors are shown in Fig.\ref{figure3}, it is evident that our proposed EaMtNet outperforms the three SOTA methods. Some quantitative analysis results are shown in Tab.\ref{tab1} and Tab.\ref{tab2}, our network achieves high performance with the DSC of 90.01$\pm$1.23\% (5.39\% higher than the second-best). The results demonstrate that edge-aware, multi-modality fusion, and uncertainty prediction are essential for segmentation.

\subsubsection{Ablation study.}  To verify the contributions of edge-aware feature aggregation (EaFA) and uncertainty, we performed ablation study and compared and performance of different networks. First, we removed the EaFA and used concatenate, meaning we removed fusion multi-modality (No-EaFA). Then, we removed the uncertainty task (No-Uncertainty). The quantitative analysis results of these ablation studies are shown in Tab.\ref{tab1}. Our method exhibits high performance in both segmentation and quantification, indicating that each component of the EaMtNet plays a vital role in liver tumor segmentation and quantification.

\subsubsection{Performance comparison with state-of-the-art.} The EaMtNet has been validated and compared with three SOTA segmentation methods and two SOTA quantification methods (i.e., ResNet-50 \cite{he2016deep} and DenseNet \cite{huang2017densely}). Furthermore, the channel number of the first convolution layer in the two quantification comparison methods is set to 2 to ensure the consistency of input modalities. The visual segmentation results are shown in Fig.3. Moreover, the quantitative results (as shown in Tab.\ref{tab2}) corresponding to the visualization results (i.e., Fig. \ref{figure3}) obtained from the existing experiments further demonstrate that our method outperforms the three SOTA methods. Specifically, compared with the second-best approach, the DSC is boosted from 84.62$\pm$1.45\% to 90.01$\pm$1.23\%. The quantitative analysis results are shown in Tab.\ref{tab3}. It is evident that our method outperforms the two SOTA methods with a large margin in all metrics, owing to the proposed multi-modality fusing and multi-task association.

 \begin{table}[t]
	\center{}
	\caption{The quantitative evaluation of segmentation. DSC is used to evaluate the performance of our EaMtNet and three SOTA methods.
	}\label{tab2}
	\begin{tabular}{|l|l|l|l|l|l|l|}
		\hline
		               & UNet &UNet++& TransUNet & Our method \\
		\hline				        
		DSC(\%) & 76.59$\pm$1.86  & 80.97$\pm$2.37  & 84.62$\pm$1.45 & 90.01$\pm$1.23  \\	           		
		
		                    		\hline
	\end{tabular}
\end{table}

 \begin{table}[t]
	\center{}
	\caption{The quantitative evaluation of the multi-index quantification. The criteria of MAE is used to evaluate the performance of our EaMtNet and two SOTA methods.
	}\label{tab3}
	\begin{tabular}{|l|l|l|l|l|l|l}
		\hline
		& MD($mm$)&X($mm$)&Y($mm$)&Area($cm^2$)\\
		\hline
		ResNet-50&6.24$\pm$2.81&3.15$\pm$1.25&3.38$\pm$1.27&31.32$\pm$8.47\\
		
		DenseNet&4.85$\pm$1.67&2.73$\pm$0.89&2.95$\pm$1.15&25.37$\pm$7.63\\
	           				   		
		Our method & 2.72$\pm$0.58&1.87$\pm$0.76&2.14$\pm$0.93&15.76$\pm$8.02\\
		                    		\hline
	\end{tabular}
\end{table}

 \section{Conclusion} 

In this paper, we have proposed an EaMtNet for the simultaneous segmentation and multi-index quantification of liver tumors on multi-modality NCMRI. The new EaFA enhances edge awareness by utilizing boundary information as prior knowledge while capturing the long-range dependency of features to improve feature selection and fusion. Additionally, multi-task leverages the prediction discrepancy to estimate uncertainty, thereby improving segmentation and quantification performance. Extensive experiments have demonstrated the proposed model outperforms the SOTA methods in terms of DSC and MAE, with great potential to be a diagnostic tool for doctors.

%\subsubsection{Acknowledgements} Please place your acknowledgments at
%the end of the paper, preceded by an unnumbered run-in heading (i.e.
%3rd-level heading).

\section{Acknowledgements.} This work is partly supported by the Natural Sciences and Engineering Research Council of Canada (NSERC) and TMU FOS Postdoctoral Fellowship.


\bibliographystyle{splncs04}
\bibliography{samplepaper}
\end{document}